\documentclass{PoS}
\usepackage{amsmath}
\usepackage{amssymb}
\def\rQCED{{\rm QCED}}
\newcommand{\FYFS}{F_{\mathrm{YFS}}}
\title{Hard Corrections in Precision QCD for LHC and FCC Physics: A New Approach}

\ShortTitle{Hard Corrections in Precision QCD for LHC and FCC Physics: A New Approach}

\author{\speaker{B.F.L. Ward}\\%
        Baylor University\\
        E-mail: \email{bfl\_ward@baylor.edu}}

\abstract{With an eye toward the usual unphysical divergence of hard fixed-order corrections in predictions for the processes probed in high energy colliding hadron beam devices as one approaches the soft limit, we present a new approach to the realization of such corrections, with some emphasis on the LHC and the future FCC devices. We show that the respective divergence is removed in our approach. This means that we would render the standard results to be closer to the observed exclusive distributions. While we stress that the approach has general applicability, we use the single $Z/\gamma^∗$ production and decay to lepton pairs as our prototypical example. Accordingly, our work opens another part of the way to rigorous baselines for the determination of the theoretical precision tags for LHC physics, with an attendant generalization to the future FCC.}

\FullConference{38th International Conference on High Energy Physics\\
		  3-10 August 2016\\
                      Chicago, USA\\
                      {\bf\rm BU-HEPP-16-04, Sep., 2016}}
\begin{document}

\section{Introduction}
In this current era of precision QCD, i.e., precision tags of 1\% or better on QCD processes, we need rigorous baselines for the attendant theoretical predictions.In our
exact amplitude resummation approach\cite{ampres1,ampres2,ampres3,ampres4} to such predictions, this amounts the construction of a semi-analytical realization of the predictions of the MC event generator used to make the predictions. This in turn requires an attendant realization of the corresponding hard gluon residuals in Refs.~\cite{ampres1,ampres2,ampres3,ampres4}
which derive from exact fixed order perturbation theory. Unfortunately, even though we have the exact NNLO result\cite{fewz} for example for the single $Z/\gamma^*$ production and decay to lepton pairs, the soft limit of the calculation is at considerable variance with the data, as one can see in Refs.~\cite{data1,data2}. It is clear that to address the precision of such results as those in \cite{fewz}, we need to tame the respective soft limits therein.\par
We can already see the problem at the NLO level. In order to properly assess the theoretical precision tag implied by our comparisons of MC@NLO\cite{mcatnlo}/Herwiri1.031\cite{ampres3} and MC@NLO/ Herwig6.5\cite{hwg65}
vs the LHC data in Refs.~\cite{ampres2,ampres3,mpla2016-am}, we need a baseline on the corresponding exact NLO results. The NLO version of the type of behavior we discussed for NNLO
exact results is illustrated in Fig. 3 in Ref.~\cite{mpla2016-bflw} and is discussed in considerable detail in Ref.~\cite{rick}, where in eq.(5.5.30) of the latter reference it is shown that
\begin{equation}
\frac{G^{DY}_{p\rightarrow q}(x,Q^2)}{G_{p\rightarrow q}(x,Q^2)}\operatornamewithlimits{\rightarrow}_{x\rightarrow 1} 1+\frac{2\alpha_s(Q^2)}{3\pi}\ln^2(1-x).
\label{rickeq1}
\end{equation}
Here, $G^{DY}_{p\rightarrow q}$ ($G_{p\rightarrow q}$) is the attendant Drell-Yan (DIS) structure function in the notation of Ref.~\cite{rick}. This behavior in (\ref{rickeq1})
is outside the scope of observable data, at the LHC or the newly conceived  FCC, and calls into question what a precision tag could even mean?\par
In what follows, we present a new approach to hard corrections in QCD which will be seen to tame the divergent behavior in (\ref{rickeq1}). The discussion will proceed as follows.
In the next section, we give a brief review of the exact amplitude-based resummation theory that we use. In Section 3, we use the exact NLO Drell-Yan formula from Refs`\cite{dy1,dy2} to compute the corresponding exact amplitude-based resummed baseline NLO formula. Section 3 also contains our summary remarks. A longer version of the material presented here has already appeared in Ref.~\cite{mpla2016-bflw}.\par
\section{Brief Review of Exact Amplitude-Based Resummation Theory}
The master formula for the exact amplitude-based resummation theory that we use is 
\begin{eqnarray}
&d\bar\sigma_{\rm res} = e^{\rm SUM_{IR}(QCED)}
   \sum_{{n,m}=0}^\infty\frac{1}{n!m!}\int\prod_{j_1=1}^n\frac{d^3k_{j_1}}{k_{j_1}} \cr
&\prod_{j_2=1}^m\frac{d^3{k'}_{j_2}}{{k'}_{j_2}}
\int\frac{d^4y}{(2\pi)^4}e^{iy\cdot(p_1+q_1-p_2-q_2-\sum k_{j_1}-\sum {k'}_{j_2})+
D_\rQCED} \cr
&\tilde{\bar\beta}_{n,m}(k_1,\ldots,k_n;k'_1,\ldots,k'_m)\frac{d^3p_2}{p_2^{\,0}}\frac{d^3q_2}{q_2^{\,0}}.
\label{subp15b}
\end{eqnarray}\noindent
Here, $d\bar\sigma_{\rm res}$ is either the reduced cross section
$d\hat\sigma_{\rm res}$ or the differential rate associated to a
DGLAP-CS~\cite{dglap,cs} kernel involved in the evolution of PDF's and 
where the {\em new} (YFS-style~\cite{yfs-jw,yfs}) {\em non-Abelian} residuals 
$\tilde{\bar\beta}_{n,m}(k_1,\ldots,k_n;k'_1,\ldots,k'_m)$ have $n$ hard gluons and $m$ hard photons and we show the final state with two hard final
partons with momenta $p_2,\; q_2$ specified for a generic $2f$ final state for
definiteness. The infrared functions ${\rm SUM_{IR}(QCED)},\; D_\rQCED\; $ are
defined in Refs.~\cite{ampres4,irdglap1,irdglap2} to which we refer the reader accordingly. We would note that, as shown in Ref.~\cite{ampres4}, the
new residuals $\tilde{\bar\beta}_{m,n}$ 
allow rigorous shower/ME matching via their shower subtracted analogs:
in (\ref{subp15b}) we make the replacements
\begin{equation}
\tilde{\bar\beta}_{n,m}\rightarrow \hat{\tilde{\bar\beta}}_{n,m}
\end{equation}
where the $\hat{\tilde{\bar\beta}}_{n,m}$ have had all effects in the showers
associated to the attendant PDF's $\{F_j\}$ removed from them. In this connection, 
we have in mind the standard formula a hard LHC(FCC) scattering process:
\begin{equation}
d\sigma =\sum_{i,j}\int dx_1dx_2F_i(x_1)F_j(x_2)d\hat\sigma_{\text{res}}(x_1x_2s),
\label{bscfrla}
\end{equation}
where  
$d\hat\sigma_{\text{res}}$ is given in (\ref{subp15b})
and thus is consistent~\cite{ampres1,ampres2,ampres3,ampres4}
with our achieving a total precision tag of 1\% or better for the total 
theoretical precision of (\ref{bscfrla}). \par
The relationship between the residuals $\hat{\tilde{\bar\beta}}_{n,m}$ and the exact NLO corrections in the MC@NLO is given in
eqs.(7-8) in Ref.~\cite{mpla2016-bflw}. From this relationship it follows that any study of the precision of results derived from (\ref{subp15b})
and used in (\ref{bscfrla}) necessarily involves the study of the precision of the corresponding NLO exact results. If we have the behavior in
(\ref{rickeq1}),  one would have to question the meaning of such a study in relation to LHC and FCC data. In the next section, we address this issue.\par
\section{Baseline Exact Amplitude-Based Resummed NLO Drell-Yan Formula}
 Focusing on the $\gamma^*$ part of $Z/\gamma^*$ with just one flavor of unit charge for reasons of pedagogy, we recall from Refs.~\cite{dy1,dy2}
the exact NLO differential cross section formula
\begin{equation}
\begin{split}
\frac{d\sigma^{DY}}{dQ^2}&=\frac{4\pi\alpha^2}{9sQ^2}\int_0^1\frac{dx_1}{x_1}\int_0^1\frac{dx_2}{x_2}\big\{\left[q^{(1)}(x_1)\bar{q}^{(2)}(x_2)+(1\leftrightarrow 2)\right]\big[\delta(1-z_{12})\\
&\quad + \alpha_s(t)\theta(1-z_{12})(\frac{1}{2\pi}P_{qq}(z_{12})(2t)+f^{DY}_q(z_{12}))\big]\\
&\quad +\left[(q^{(1)}(x_1)+\bar{q}^{(1)}(x_1))G^{(2)}(x_2)+(1\leftrightarrow 2)\right] \\
&\quad \times [\alpha_s(t)\theta(1-z_{12})(\frac{1}{2\pi}P_{qG}(z_{12})t+f^{DY}_G(z_{12}))]\big\}
\end{split}
\label{guido-eq2}
\end{equation}
where $z_{12}=\tau/(x_1x_2), \; \tau=Q^2/s$ in the usual conventions~\cite{rick,dy1,dy2}, the labels $1$ and $2$ refer to the two respective incoming protons
and we follow the generic notation of Refs. ~\cite{rick,dy1} here. In (\ref{guido-eq2}), unimproved DGLAP-CS~\cite{dglap,cs} kernels are
\begin{align}
P_{qq}(z)&= C_F \left[\frac{1+z^2}{(1-z)_+} + \frac{3}{2}\delta(1-z)\right],\nonumber\\
P_{qG}(z)&=\frac{1}{2}(z^2+(1-z)^2),
\label{guido-eq3}
\end{align} 
where we define $t=\ln(Q^2/\mu^2)$ following Refs.\cite{dy1,rick} so that
$\mu$ is the 't Hooft~\cite{thft-mass} unity of mass.
The scheme dependent hard correction terms are given as follows~\cite{dy1,dy2} if one uses massless quarks and gluons and dimensional regularization, for example:
\begin{equation}
\begin{split}
\alpha_s f^{DY}_G(z)&=\frac{\alpha_s}{2\pi}\frac{1}{2}[(z^2+(1-z)^2)\ln\frac{(1-z)^2}{z}-\frac{3}{2}z^2+z+\frac{3}{2}+2P_{qG}(z)\zeta]\\
\alpha_s f^{DY}_q(z)&=C_F\frac{\alpha_s}{2\pi}\Big[4(1+z^2)\left(\frac{\ln(1-z)}{1-z}\right)_+ -2\frac{1+z^2}{1-z}\ln{z}\\
& \quad+\left(\frac{2\pi^2}{3}-8\right)\delta(1-z)+\frac{2}{C_F}P_{qq}(z)\zeta\Big] 
\end{split}
\label{guido-eq4}
\end{equation}
where we define~\cite{dy2} $\zeta=-\frac{1}{\epsilon}+C_E-\ln{4\pi}$ for 
$\epsilon=2-n/2$ when $n$ is the dimension of space-time. $C_E$ is Euler-Mascheroni constant.
In the $\overline{\text{MS}}$ scheme, the terms proportional to $\zeta$ are removed 
by mass factorization, which also replaces $\mu$ by $\Lambda$ in $t$ following Ref.~\cite{rick}. This leaves the +-functions in the hard corrections and it is the divergent behavior of these distributions as $z\rightarrow 1$ that produces the attendant unphysical results referenced above. 
.\par
To fix this, we imbed the calculation of the hard correction terms into the master formula
(\ref{subp15b}) restricted to its QCD aspect. This gives the following resummed
version of (\ref{guido-eq2}):
\begin{equation}
\begin{split}
\frac{d\sigma^{DY}_{res}}{dQ^2}&=\frac{4\pi\alpha^2}{9sQ^2}\int_0^1\frac{dx_1}{x_1}\int_0^1\frac{dx_2}{x_2}\big\{\left[q^{(1)}(x_1)\bar{q}^{(2)}(x_2)+(1\leftrightarrow 2)\right]2\gamma_qF_{YFS}(2\gamma_q)(1-z_{12})^{2\gamma_q-1}e^{\delta_q}\\
&\quad \times \theta(1-z_{12})\big[ 1+\gamma_q -7C_F\frac{\alpha_s}{2\pi}+ (1-z_{12})(-1+\frac{1-z_{12}}{2})\\
&\quad +2\gamma_q(-\frac{1-z_{12}}{2}-\frac{z_{12}^2}{4}\ln{z_{12}})\\
&\quad + \alpha_s(t)\frac{(1-z_{12})}{2\gamma_q}f^{DY}_q(z_{12})\big]\\
&\quad +\left[(q^{(1)}(x_1)+\bar{q}^{(1)}(x_1))G^{(2)}(x_2)+(1\leftrightarrow 2)\right] \\
&\quad \times \gamma_G F_{YFS}(\gamma_G)e^{\frac{\delta_G}{2}}[\alpha_s(t)\theta(1-z_{12})\big(\frac{t}{2\pi\gamma_G}(\frac{1}{2}(z_{12}^2(1-z_{12})^{\gamma_G}+(1-z_{12})^2z_{12}^{\gamma_G}))\\
&\quad +f^{DY'}_G(z_{12})/\gamma_G\big)]\big\}
\end{split}
\label{guido-eq5}
\end{equation}
where we have introduced here
\begin{equation}
\begin{split}
\alpha_s f^{DY'}_G(z)&=\frac{\alpha_s}{2\pi}\frac{1}{2}[(z^2(1-z)^{\gamma_G}+(1-z)^2z^{\gamma_G})\ln\frac{(1-z)^2}{z}-\frac{3}{2}z^2(1-z)^{\gamma_G}+z(1-z)^{\gamma_G}\\
&\quad +\frac{3}{4}((1-z)^{\gamma_G}+z^{\gamma_G})],\\
\end{split}
\label{guido-eq6}
\end{equation}
and the following exponents and YFS infrared function, $\FYFS$, already needed for the IR-improvement of DGLAP-CS theory in Refs.~\cite{irdglap1,irdglap2}:
\begin{align}
\gamma_q &= C_F\frac{\alpha_s}{\pi}t=\frac{4C_F}{\beta_0}, \qquad \qquad
\delta_q =\frac{\gamma_q}{2}+\frac{\alpha_sC_F}{\pi}(\frac{\pi^2}{3}-\frac{1}{2}),\nonumber\\
\gamma_G &= C_G\frac{\alpha_s}{\pi}t=\frac{4C_G}{\beta_0}, \qquad \qquad
\delta_G =\frac{\gamma_G}{2}+\frac{\alpha_sC_G}{\pi}(\frac{\pi^2}{3}-\frac{1}{2}),\nonumber\\
\FYFS(\gamma)&=\frac{e^{-{C_E}\gamma}}{\Gamma(1+\gamma)}.
\label{resfn1}
\end{align}
We define $\beta_0=11-\frac{2}{3}n_f$ for $n_f$ active flavors 
in a standard way and $\Gamma(w)$ is Euler's gamma function of the complex variable $w$.
We have mass factorized in (\ref{guido-eq5}) and (\ref{guido-eq6}) 
as indicated above.
We see that the regime at $z_{12}\rightarrow 1$ is now under control in (\ref{guido-eq5}) so that we will no longer have the unphysical
behavior discussed above. This is the main result of this paper.\par
More precisiely, in lieu of the result in (\ref{rickeq1}), we get the
behavior such that the $\ln^2(1-x)$ on the RHS of (\ref{rickeq1}) is replaced by
$$\frac{2(1-x)^{\gamma_q}\ln(1-x)}{\gamma_q} - \frac{2(1-x)^{\gamma_q}}{\gamma_q^2},$$
and this vanishes for $x\rightarrow 1$. Our result
means that the hard correction now has the possibility to be compared rigorously
to the data in an exclusive manner. We take up such matters elsewhere.~\cite{elswh}.\par
We note that the parton shower/ME matching formulas in MC@NLO and in POWHEG~\cite{powheg} do not remove the IR divergence which we just tamed -- the latter retains the NLO correction with its bad IR limit in the soft regime for $z_{12} \rightarrow 1$ and the former replaces the bad IR behavior of the NLO correction with that of the unimproved parton shower real emission at the same order which is infrared divergent for $z_{12} \rightarrow 1$ and requires an ad hoc IR cut-off $k_0$-parameter, as we have discussed in Ref.~\cite{ampres3}. {\em No such parameter is needed in our new approach.}\par
In summary, we have introduced a new approach to hard corrections in perturbative QCD which allows the realization of the same type of semi-analytical baselines for QCD that we and our collaborators had in Refs.~\cite{yfs-jw} for the higher order 
corrections in the 
Standard Model EW theory.
We look with excitement to its exploitation in precision LHC and FCC physics scenarios.
In closing, we 
thank Prof. Ignatios Antoniadis for the support and kind 
hospitality of the CERN TH Unit while part of this work was completed.\par

\end{document}